\documentstyle[preprint,aps,graphicx,eqsecnum]{revtex}

\makeatletter

\def\compoundrel#1\over#2{\mathpalette\compoundreL{{#1}\over{#2}}}
\def\compoundreL#1#2{\compoundREL#1#2}
\def\compoundREL#1#2\over#3{\mathrel
  {\vcenter{\hbox{$\m@th\buildrel{#1#2}\over{#1#3}$}}}}
\makeatother
\newcommand{\be}{\begin{equation}}
\newcommand{\ee}{\end{equation}}
\newcommand{\bea}{\begin{eqnarray}}
\newcommand{\eea}{\end{eqnarray}}
\newcommand{\bref}[1]{(\ref{#1})}
\newcommand{\mapright}[1]{%
	\smash{\mathop{%
	\hbox to 1cm{\rightarrowfill}}\limits^{#1}}}

\begin{document}
\draft
\title{{Universal Texture of Quark and Lepton  Mass Matrices \\
and a Discrete Symmetry Z$_3$}}

\author{Yoshio KOIDE\thanks{
On leave at CERN, Geneva, Switzerland.
}}
\address{
Department of Physics, 
University of Shizuoka, 52-1 Yada, Shizuoka, 422-8526 Japan}
\author{Hiroyuki NISHIURA}
\address{
Department of General Education, 
Junior College of Osaka Institute of Technology, \\
Asahi-ku, Osaka,535-8585 Japan}
\author{Koichi MATSUDA, Tatsuru KIKUCHI, and Takeshi FUKUYAMA}
\address{
Department of Physics, 
Ritsumeikan University, Kusatsu, Shiga, 525-8577 Japan}

%\date{\today}
\maketitle

\begin{abstract}

Recent neutrino data have been favourable to a nearly bimaximal mixing, 
which suggests a simple form of the neutrino mass matrix.
Stimulated by this matrix form,  a possibility that all the mass 
matrices of quarks and leptons have the same form as in the 
neutrinos is investigated.
The mass matrix form is constrained by a discrete symmetry Z$_3$ and
a permutation symmetry S$_2$.
The model, of course, leads to a nearly bimaximal mixing 
for the lepton sectors, while, for the quark sectors, it can lead to 
reasonable values of the CKM mixing matrix and masses.
\end{abstract}

%%%%%%%%%%%%%%%%%%%%%%%%%%%%%%%%%%%%%%%%%%%%%%%%%%%%%%%%%%%%%%%%%%%%%%
\section{Introduction}

Recent neutrino oscillation experiments\cite{skamioka} have highly
suggested a nearly bimaximal mixing $(\sin^2 2\theta_{12}\sim 1$, 
$\sin^2 2\theta_{23}\simeq 1)$ together with a small ratio 
$R\equiv \Delta m^2_{12}/\Delta m^2_{23} \sim 10^{-2}$.
This can be explained by assuming a neutrino mass matrix form
\cite{Fukuyama}--\cite{Nishiura}
with a permutation symmetry between second and third generations.
We think that quarks and leptons should be unified.
It is therefore interesting to investigate a possibility that all
the mass matrices of the quarks and leptons have the same matrix form,
which leads to a nearly bimaximal mixing and $U_{13}=0$ in the neutrino
sector, against the conventional picture that the mass matrix forms in the
quark sectors will take somewhat different structures from those in the 
lepton sectors.
In the present paper, we will assume that the mass matrix form is invariant
under a discrete symmetry Z$_3$ and a permutation symmetry S$_2$.

Phenomenologically, our mass matrices 
\(M_u\), \(M_d\), \(M_\nu\) and  \(M_e\) 
(mass matrices of up quarks (\(u,c,t\)), down quarks (\(d,s,b\)), 
neutrinos (\(\nu_e,\nu_\mu,\nu_\tau\)) and 
charged leptons (\(e,\mu,\tau\)), 
respectively) are given as follows:
\begin{equation}
M_f = P_{Lf}^\dagger \widehat{M}_f P_{Rf}, \ 
\quad \quad \quad \quad 
\end{equation}
with
\begin{equation}
\widehat{M}_f=
\left(
\begin{array}{lll}
\ 0 & \ A_f & \ A_f \\
\ A_f & \ B_f & \ C_f \\
\ A_f & \ C_f & \ B_f \\
\end{array}
\right) \ \, \left(f=u,d,\nu,e\right),
\label{texture}
\end{equation}
where $P_{Lf}$ and $P_{Rf}$ are the diagonal phase matrices 
and $A_f$, $B_f$, and $C_f$ are real parameters. 
Namely the components are different in $\widehat{M}_f$, 
but their mutual relations are the same.
This structure of mass matrix was previously suggested and used 
for the neutrino mass matrix in Refs\cite{Fukuyama}--\cite{Nishiura},  
using the basis where the charged-lepton mass matrix is diagonal, 
motivated by the experimental finding of maximal \(\nu_\mu\)--\(\nu_\tau\) 
mixing\cite{skamioka}.  In this paper, we consider that this structure 
is fundamental for both quarks and leptons, although it was speculated 
from the neutrino sector. Therefore, we assume that all the mass matrices 
have this structure.

Let us look at the universal characters of the model.  
Hereafter, for brevity, we will omit the flavour index.
The eigen-masses $m_i$ of Eq. \bref{texture} are given by
\begin{eqnarray}
-m_1& =&
\frac{1}{2}
\left(B+C-\sqrt{8A^2 + (B+C)^2}
\right) ,\\
m_2& =&\frac{1}{2}
\left(B+C+\sqrt{8A^2 + (B+C)^2}
\right) ,\\
m_3& =&B-C. 
\end{eqnarray}
The texture's components of \(\widehat{M}\) are expressed 
in terms of eigen-masses $m_i$ as 
\begin{eqnarray}
A & =&\sqrt{\frac{m_2m_1}{2}}  ,\nonumber\\
B & =&\frac{1}{2}m_3 \left(1+\frac{m_2-m_1}{m_3}\right) ,\label{eq2003}\\
C & =&-\frac{1}{2}m_3\left(1-\frac{m_2-m_1}{m_3}\right) .\nonumber 
\end{eqnarray}
That is, 
\(\widehat{M}\) is diagonalized by an orthogonal matrix \(O\) as
\begin{equation}
O^T\widehat{M}O=
\left(
	\begin{array}{ccc}
	-m_1 & 0 & 0\\
	0 & m_2 & 0\\
	0 &  0 & m_3
	\end{array}
\right),
\end{equation}
with
\begin{equation}
O\equiv
\left(
\begin{array}{ccc}
{ c}&
{ s}&
{0} \\
{-\frac{s}{\sqrt{2}}}&
{\frac{c}{\sqrt{2}}}&
{-\frac{1}{\sqrt{2}}} \\
{-\frac{s}{\sqrt{2}}}&
{\frac{c}{\sqrt{2}}}&
{\frac{1}{\sqrt{2}}}
\end{array}
\right). \label{eq990114} \\
\label{O}
\end{equation}
Here \(c\) and \(s\) are defined by 
\begin{equation}
c=\sqrt{\frac{m_2}{m_2+m_1}},\quad
s=\sqrt{\frac{m_1}{m_2+m_1}}.\label{eq2012}
\end{equation}
It should be noted that the elements of \(O\) are independent of \(m_3\) 
because of the above structure of \(\widehat{M}\).

The zeros in this mass matrix are constrained by the discrete 
symmetry that is discussed in the next section, defined at 
a unification scale (the scale does not always mean ``grand
unification scale"). This discrete symmetry is  
broken below $\mu=M_R$, at which the right-handed neutrinos acquire
heavy Majorana masses, as we discuss in Sec.~IV.
Therefore, the matrix form (1.1) will, in general, be  changed by
renormalization group equation (RGE) effects.
Nevertheless, we would like to emphasize that
we can use the expression (1.1) with (1.2) for the predictions 
of the physical quantities in the low-energy region.
This will be discussed in the appendix.

This article is organized as follows. 
In Sec.~II we discuss the symmetry property of our model.
Our model is realized when we consider two Higgs doublets in each 
up-type and down-type quark (lepton) mass matrices. 
The quark mixing matrix in the present model is argued in Sec.~III. 
In Sec.~IV, the lepton mixing matrix is analyzed.
Sec.~V is devoted to a summary.

%%%%%%%%%%  chap 2  %%%%%%%%%%%%%%
\section{Z$_3$ symmetry and mass matrix form}

We assume a permutation symmetry between second and third generations,
except for the phase factors.
However, the condition $(\widehat{M}_f)_{11}=0$ cannot be
derived from such a symmetry.
Therefore, in addition to the $2 \leftrightarrow 3$ symmetry,
we assume a discrete symmetry Z$_3$, 
under which symmetry the quark and lepton fields $\psi_L$,
which belong to $10_L$, $\overline{5}_L$ and $1_L$ of SU(5) 
($1_L={\nu}_R^c$), are transformed as
\be
\begin{array}{c}
\psi_{1L} \rightarrow \psi_{1L} , \\
\psi_{2L} \rightarrow \omega \psi_{2L} , \\
\psi_{3L} \rightarrow \omega \psi_{3L}, \\
\end{array}
\ee
where $\omega^3=+1$.
(Although we use a terminology of SU(5), at present,
we do not consider the SU(5) grand unification.)
Then, the bilinear terms $\overline{q}_{Li} u_{Rj}$, 
$\overline{q}_{Li} d_{Rj}$, $\overline{\ell}_{Li} \nu_{Rj}$, 
$\overline{\ell}_{Li} e_{Rj}$ and $\overline{\nu}_{Ri}^c \nu_{Rj}$
[$\nu_R^c =(\nu_R)^c =C \overline{\nu_R}^T$ and
$\overline{\nu}_R^c =\overline{(\nu_R^c)}$] 
are transformed as follows:
\begin{equation}
\left( 
\begin{array}{ccc}
1 & \omega^2 & \omega^2 \\
\omega^2 & \omega & \omega \\
\omega^2 & \omega & \omega \\
\end{array} \right) \ ,
\end{equation} 
where
\begin{equation}
q_{L} =\left(
\begin{array}{c}
u_L \\
d_L
\end{array} \right) \ , \ \ \ 
\ell_{L} =\left(
\begin{array}{c}
\nu_L \\
e^-_L
\end{array} \right) \ .
\end{equation}
Therefore, if we assume two SU(2) doublet Higgs scalars $H_1$ and $H_2$, 
which are transformed as
\be
H_1 \rightarrow \omega H_1 , \ \ \ H_2 \rightarrow \omega^2 H_2 , 
\ee
the Yukawa interactions are given  as follows
\begin{eqnarray}
H_{{\rm int}} &=&
 \sum_{A=1,2}\left( Y^{u}_{(A)ij}\overline{q}_{Li}\widetilde{H}_{A} u_{Rj}
+ Y^{d}_{(A)ij}\overline{q}_{Li}{H}_A d_{Rj} \right)\nonumber \\
&+&  \sum_{A=1,2}\left( Y^{\nu}_{(A)ij}\overline{\ell}_{Li}\widetilde{H}_{A} 
\nu_{Rj}
+ Y^{e}_{(A)ij}\overline{\ell}_{Li}{H}_A e_{Rj} \right) \\
&+& \left( Y^{R}_{(1)ij}\overline{\nu}_{Ri}^c \widetilde{\Phi}^0 \nu_{Rj} 
   + Y^{R}_{(2)ij}\overline{\nu}_{Ri}^c {\Phi}^0 \nu_{Rj} \right)
+ {\rm h.c.} \ ,\nonumber
\end{eqnarray}
where
\begin{equation}
H_A =\left(
\begin{array}{c}
H^+_A \\
H^0_A
\end{array} \right) \ , \ \ \ 
\widetilde{H}_A =\left(
\begin{array}{c}
\overline{H}^0_A \\
-H^-_A
\end{array} \right) \ , 
%\ \ \
%\widehat{\Phi}^0 = \overline{\Phi^0} \ ,  
\label{tilde}
\end{equation}
so that
\be
Y^u_{(1)},\ Y^d_{(2)},\ Y^\nu_{(1)},\ Y^e_{(2)},\ Y^R_{(2)} =
\left(
\begin{array}{ccc}
0 & 0 & 0 \\
0 & \ast & \ast \\
0 & \ast & \ast \\
\end{array}
\right)
\ , \ \ \ \ 
Y^u_{(2)},\ Y^d_{(1)},\ Y^\nu_{(2)},\ Y^e_{(1)},\ Y^R_{(1)} =
\left(
\begin{array}{ccc}
0 & \ast & \ast \\ 
\ast & 0 & 0 \\ 
\ast & 0 & 0 \\
\end{array}
\right).
\ee
In (2.7), the symbol $*$ denotes  non-zero quantities. 
Here, in order to give  heavy Majorana masses of the right-handed neutrinos
$\nu_R$, we have assumed an SU(2) singlet Higgs scalar $\Phi^0$, which is 
transformed  as $H_1$.

In the present model, the phase difference
${\rm arg}(Y_{(1)}^f +Y_{(2)}^f)_{21}-{\rm arg}(Y_{(1)}^f +Y_{(2)}^f)_{31}$
plays an essential role.
Therefore, for the permutation symmetry S$_2$, we put the following
assumption: the permutation symmetry can be applied
to only the special basis that the all Yukawa coupling constants are
real. (Of course, for the Z$_3$ symmetry, such an assumption is not
required.)
We consider that the phase factors are caused by an additional
mechanism after the requirement of the permutation symmetry S$_2$
(after the manifestation of the linear combination $Y_{(1)}+ 
Y_{(2)}$).
In the present paper, we consider that although the Z$_3$ symmetry
is rigorously defined for the fields by (2.1), the permutation
symmetry S$_2$ is rather phenomenological one (i.e. Ansatz) for
the mass matrix shape.
Then, under such the S$_2$ symmetry, 
the general forms of $Y_f \equiv Y_{(1)}^f +Y_{(2)}^f$ are given by 
\be
Y_f=P_{Lf}^\dagger \widehat{Y}_f P_{Rf} =
\left(
\begin{array}{ccc}
0 & a e^{-i(\delta_{L1}^f-\delta_{R2}^f)} & 
a e^{-i(\delta_{L1}^f-\delta_{R3}^f)} \\
a e^{-i(\delta_{L2}^f-\delta_{R1}^f)}  &
 b e^{-i(\delta_{L2}^f-\delta_{R2}^f)}  & 
c e^{-i(\delta_{L2}^f-\delta_{R3}^f)}  \\
a e^{-i(\delta_{L3}^f-\delta_{R1}^f)}  & 
c e^{-i(\delta_{L3}^f-\delta_{R2}^f)}  & 
b e^{-i(\delta_{L3}^f-\delta_{R3}^f)}  \\
\end{array}
\right).
\ee

We have already assumed that $\psi_L=(\nu_,e_L,d_R^c;
u_L, d_L, u_R^c, e_R^c; \nu_R^c)$ have the same
transformation (2.1) under the discrete symmetry Z$_3$,
so that $\overline{\psi}_{Li} \psi_{Lj}^c$ are transformed
as (2.2).
{}From this analogy, we assume that the phase matrices
$P_{Lf}$ and $P_{Rf}$ come from the replacement 
$\psi_L \rightarrow P_f \psi_L$, i.e.
\begin{equation}
\overline{\psi}_L Y_f \psi_L^c \rightarrow 
\overline{\psi}_L P_f^\dagger \widehat{Y}_f P_f^\dagger \psi_L^c \ .
\end{equation}
However, differently from the transformation (2.1),
we do not assume in (2.9) that all the phase matrices $P_f$
are identical, but we assume that they are flavour dependent.
This explains the assumption
\begin{equation}
\delta_{Li}^f=-\delta_{Ri}^f \equiv \delta_i^f \ ,
\end{equation}
in the expression (2.8).
(However, this assumption (2.10) is not essential for the
numerical predictions in the present paper, because the
predictions of the physical quantities depend on only 
the phases $\delta_{Li}^f$.)

Since the present model has two Higgs doublets horizontally, in general,
flavour-changing neutral currents (FCNCs) are caused by
the exchange of Higgs scalars.
However, this FCNC problem is a common subject 
to be overcome not only in the present model but also in most 
models with two Higgs doublets. 
The conventional mass matrix models 
based on a GUT scenario cannot give realistic mass 
matrices without assuming more than two Higgs scalars
\cite{howmanyH}.
Besides, if we admit that two such  scalars remain until
the low energy scale, the well-known beautiful coincidence
of the gauge coupling constants at $\mu \sim 10^{16}$ GeV
will be spoiled.
Although the present model is not based on a GUT scenario,
as are the conventional mass matrix models, for the FCNC problem, 
we optimistically consider 
that only one component of the linear combinations 
among those Higgs scalars survives at the low energy scale 
$\mu=m_Z$, while the other component is decoupled at
$\mu < M_X$ \cite{Fuk-Okd}.
The study of the RGE effects given in the appendix will be 
based on such an ``effective" one-Higgs scalar scenario.

\vspace{3mm}

%%%%%%%%%   chap 3 %%%%%%%%%%%%%%%
\section{quark mixing matrix}

The quark mass matrices 
\begin{equation}
M_f = P_{f}^\dagger\widehat{M}_fP_{f}^\dagger  \ \ (f=u,d),
\end{equation}
are diagonalized by the bi-unitary transformation
\begin{equation}
D_f  =  U_{Lf}^\dagger M_f U_{Rf}\ , 
\end{equation}
where $U_{Lf}\equiv P_f^\dagger O_f$, $U_{Rf}\equiv P_f O_f$, 
and $O_d$ ($O_u$) is given by Eq.~(\ref{O}). 
Then, the Cabibbo--Kobayashi--Maskawa (CKM) \cite{CKM} quark mixing 
matrix \(V\) is given by
\begin{eqnarray}
V&=&U^\dagger_{Lu}U_{Ld}=O^T_uP_{u}P^\dagger_{d} O_d\nonumber\\[.1in]
& =&
\left(
\begin{array}{ccc}
c_uc_d+\rho s_u s_d \quad & c_u s_d-\rho s_u c_d \quad & -{\sigma}s_u \\
s_uc_d-\rho c_u s_d \quad & s_u s_d+\rho c_u c_d \quad & {\sigma}c_u \\
-{\sigma}s_d \quad & {\sigma}c_d \quad & \rho \\
\end{array}
\right),\label{eq-ourckm} 
\end{eqnarray}
where  \(\rho\) and \(\sigma\) are defined by 
\begin{equation}
\rho=\frac{1}{2}(e^{i\delta_3}+e^{i\delta_2})
=\cos\frac{\delta_3 - \delta_2}{2} \exp i
\left( \frac{\delta_3 + \delta_2}{2} \right) \ ,
\end{equation}
\begin{equation}
\sigma=\frac{1}{2}(e^{i\delta_3}-e^{i\delta_2})
= \sin\frac{\delta_3 - \delta_2}{2} 
\exp i \left( \frac{\delta_3 + \delta_2}{2}+ \frac{\pi}{2}
\right) \ . 
\end{equation}
Here we have put \(P \equiv P_{u}P^\dagger_{d} \equiv 
\mbox{diag}(e^{i\delta_1}, e^{i\delta_2},e^{i\delta_3})\), and
we have taken \(\delta_1=0\) without 
loss of generality.

Then, the explicit magnitudes of the components of \(V\) are expressed as
\begin{eqnarray}
\left|V_{cb}\right| & =&\left|\sigma\right| c_u 
= \frac{\sin\frac{\delta_3-\delta_2}{2}}{\sqrt{1+m_u/m_c}},\label{eq3021}\\
\left|V_{ub}\right|& =&\left|\sigma\right| s_u 
= \frac{\sin\frac{\delta_3-\delta_2}{2}}{\sqrt{1+m_u/m_c}}\sqrt{
\frac{m_u}{m_c}},\label{eq3022}\\
\left|V_{ts}\right|& =&\left|\sigma\right| c_d 
= \frac{\sin\frac{\delta_3-\delta_2}{2}}{\sqrt{1+m_d/m_s}},\label{eq3023}\\
\left|V_{td}\right|& =&\left|\sigma\right| s_d 
= \frac{\sin\frac{\delta_3-\delta_2}{2}}{\sqrt{1+m_d/m_s}}\sqrt{
\frac{m_d}{m_s}},\label{eq3024}\\
\left| V_{us}\right|
& =&c_u s_d\left|1 -\rho{\frac{s_u}{c_u}}{\frac{c_d}{s_d}}\right|
=\sqrt{\frac{m_c}{m_c+m_u}} \sqrt{\frac{m_d}{m_s+m_d}}\nonumber \\
& \times&
\left[1 -2\cos\frac{\delta_3-\delta_2}{2}
\cos\frac{\delta_3+\delta_2}{2}
 \sqrt{\frac{m_u m_s}{m_c m_d}}
+ \cos^2 \frac{\delta_3 - \delta_2}{2}
\left( \frac{m_u m_s}{m_c m_d}\right)\right]^{\frac{1}{2}},\label{eq3025}\\
\left| V_{cd}\right|
& =&c_u s_d\left|\rho -{\frac{s_u}{c_u}}{\frac{c_d}{s_d}}\right|
=\sqrt{\frac{m_c}{m_c+m_u}} \sqrt{\frac{m_d}{m_s+m_d}}\nonumber \\
& \times&
\left[\cos^2 \frac{\delta_3 - \delta_2}{2} -2\cos\frac{\delta_3-\delta_2}{2}
\cos\frac{\delta_3+\delta_2}{2} \sqrt{\frac{m_u m_s}{m_c m_d}}
+ 
\left( \frac{m_u m_s}{m_c m_d}\right) 
\right]^{\frac{1}{2}}.\label{eq3026}
\end{eqnarray}
It should be noted that the elements of \(V\) are independent of 
\(m_t\) and \(m_b\).
The independent parameters in the expression $|V_{ij}|$ are 
$\theta_u = \tan^{-1}(m_u/m_c)$,
$\theta_d = \tan^{-1} (m_d/m_s)$, $\delta_3$, and $\delta_2$. 
Among them, the two parameters $\theta_u$ and $\theta_d$ are already 
fixed by the quark masses of the first and second generations. 
Therefore, the present model has two adjustable parameters  
$\delta_3$ and $\delta_2$ to reproduce the observed CKM matrix 
parameters \cite{PDG}:
\begin{eqnarray}
&&|V_{us}|_{\mbox{{\tiny exp}}}= 0.2196 \pm 0.0026, \quad 
|V_{cb}|_{\mbox{{\tiny exp}}}= 0.0412 \pm 0.0020, \nonumber\\
&&|V_{ub}|_{\mbox{{\tiny exp}}}= (3.6 \pm 0.7)\times 10^{-3}, \quad
\end{eqnarray}

It should be noted that the predictions
\begin{eqnarray}
\frac{\left|V_{ub}\right|}{\left|V_{cb}\right|}& =&\frac{s_u}{c_u}
=\sqrt{\frac{m_u}{m_c}}
=\sqrt{\frac{2.33}{677}}=0.0586\pm 0.0064 \ , \\
\frac{\left|V_{td}\right|}{\left|V_{ts}\right|}& =&\frac{s_d}{c_d}
=\sqrt{\frac{m_d}{m_s}}
=\sqrt{\frac{4.69}{93.4}}=0.224\pm 0.014 \ 
\end{eqnarray}
are almost independent of the RGE effects, because they do not
contain the phase difference, $(\delta_3 -\delta_2)$, which
is highly dependent on the energy scale as we discuss in the appendix
[see (A.9)] and we know that the ratios
$m_u/m_c$ and $m_d/m_s$ are almost independent of the RGE effects.
In the numerical results of (3.13) and (3.14), 
we have used the running quark mass at \(\mu=m_Z\) \cite{Fusaoka}:
\begin{equation}
\begin{array}{lll}
m_u(m_Z)=2.33^{+0.42}_{-0.45}\, \mbox{MeV},& 
m_c(m_Z)=677^{+56}_{-61}\, \mbox{MeV},\\
m_d(m_Z)=4.69^{+0.60}_{-0.66}\, \mbox{MeV},& 
m_s(m_Z)=93.4^{+11.8}_{-13.0}\, \mbox{MeV} .
\end{array}
\label{eq123103}
\end{equation}
The predicted value (3.13) is somewhat small with respect to
the present experimental value $|V_{ub}|/|V_{cb}|=0.08 \pm 0.02$,
but it is within the error.

The heavy-quark-mass-independent predictions (3.13) and
(3.14) have first been derived from a special ansatz for
quark mixings by Branco and Lavoura \cite{Branco},
and later, a similar formulation has also be given
by Fritzsch and Xing \cite{Fritzsch}.
For example, the CKM matrix $V$ is given by the form
$V=R_{12}(\theta_u) R_{23}(\theta_Q,\phi_Q) R_{12}^T(\theta_d)$
in the Fritzsch--Xing ansatz, and their rotation
$R_{23}(\theta_Q,\phi_Q)$ with a phase $\phi_Q$ corresponds
to $R_{23}(-\pi/4) P_u P_d^\dagger R_{23}^T(-\pi/4)$ in the
present model, because the present rotation given in (1.8)
is expressed as $O_f =R_{23}(-\pi/4) R_{12}(\theta_f)$.
However, we would like to emphasize that the 
$2\leftrightarrow 3$ mixing in $V$ comes from only the
relative phase difference $(\delta_2-\delta_3)$, and it is
independent of the forms of the up- and down-mixing
matrices (1.8).
The present mass matrix texture is completely different from
theirs.
The rederivation of (3.13) and (3.14) in the present model 
will illuminate the farsighted instates by Branco and Lavoura.

%\par

Next let us fix the parameters $\delta_3$ and $\delta_2$.
When we use the expressions (3.6)--(3.11)
at $\mu=m_Z$, the parameters $\delta_2$ and $\delta_3$ do not
mean the phases that are evolved from those at $\mu=M_X$.
Hereafter, we use the parameters $\delta_2$ and $\delta_3$ 
as phenomenological parameters that approximately satisfy
the relations (3.6)--(3.11) at $\mu=m_Z$.
In order to fix the value of ${\delta_3-\delta_2}$, we use the
relation (3.6), which leads to
\begin{equation}
\sin\frac{\delta_3-\delta_2}{2} = \sqrt{1+\frac{m_u}{m_c}}\left|V_{cb}
\right|_{\mbox{{\tiny exp}}}=0.0401\pm 0.0018 \ ,
\end{equation}
\begin{equation}
\delta_3 -\delta_2 = 4.59^\circ\pm 0.21^\circ. \label{eq3032}
\end{equation} 
Then, we obtain 
\begin{eqnarray}
|V_{ub}|& =&\sqrt{\frac{m_u}{m_c}}\left|V_{cb}\right|_{\mbox{{\tiny exp}}}
=0.00234\pm 0.00028,\label{eq3033}\\
|V_{ts}|& =&\sqrt{\frac{1+\frac{m_u}{m_c}}{1+\frac{m_d}{m_s}}}\left|V_{cb}
\right|_{\mbox{{\tiny exp}}}
=0.0391\pm0.0018,\label{eq3034}\\
|V_{td}|& =&\sqrt{\frac{1+\frac{m_u}{m_c}}{1+\frac{m_d}{m_s}}}
\sqrt{\frac{m_d}{m_s}}\left|V_{cb}\right|_{\mbox{{\tiny exp}}}=
0.00880\pm 0.00094,\label{eq3035}
\end{eqnarray}
which are consistent with the present experimental data.
Therefore, the value (3.17) is acceptable as reasonable.
Then, by using the value (3.17) and the expression (3.10),  
we can obtain the remaining parameter $(\delta_3+\delta_2)$ :
\begin{equation}
\delta_3+ \delta_2 = 93^\circ \pm 22^\circ \
\mbox{  or  } -80^\circ \pm 22^\circ  \ . \label{eq3036}
\end{equation}
Since $\sin(\delta_3-\delta_2)/2 \simeq 0.04$  and 
$\cos(\delta_3+\delta_2)/2 \simeq 0.2$, the present model also 
predicts the following approximated relations 
\begin{eqnarray}
\left| V_{us}\right|
& =&c_u s_d\left|1 -\rho{\frac{s_u}{c_u}}{\frac{c_d}{s_d}}\right|
\simeq\sqrt{\frac{m_d}{m_s}},\\
\left| V_{cd}\right|
& =&c_u s_d\left|\rho -{\frac{s_u}{c_u}}{\frac{c_d}{s_d}}\right|
\simeq\sqrt{\frac{m_d}{m_s}},\\
\left| V_{td}\right| & =& \left| {\sigma} \right| s_d
=\sqrt{\left|V_{cb}\right|^2+\left|V_{ub}\right|^2}
\sqrt{\frac{m_d}{m_s+m_d}} 
\simeq \left|V_{cb}\right|{\cdot} \left|V_{us}\right|.\label{eq3039}
\end{eqnarray}
\par
Using the rephasing of the up-type and down-type quarks, 
Eq.~(\ref{eq-ourckm}) is changed to the standard representation 
of the CKM quark mixing matrix  
\begin{eqnarray}
V_{\rm std} &=& \mbox{diag}(e^{\alpha_1^u},e^{\alpha_2^u},e^{\alpha_2^u})  \ V \ 
\mbox{diag}(e^{\alpha_1^d},e^{\alpha_2^d},e^{\alpha_2^d}) \nonumber \\
&=&
\left(
\begin{array}{ccc}
c_{13}c_{12} & c_{13}s_{12} & s_{13}e^{-i\delta} \\
-c_{23}s_{12}-s_{23}c_{12}s_{13} e^{i\delta}
&c_{23}c_{12}-s_{23}s_{12}s_{13} e^{i\delta} 
&s_{23}c_{13} \\
s_{23}s_{12}-c_{23}c_{12}s_{13} e^{i\delta}
 & -s_{23}c_{12}-c_{23}s_{12}s_{13} e^{i\delta} 
& c_{23}c_{13} \\
\end{array}
\right) \ .
\label{stdrep}
\end{eqnarray}
Here, \(\alpha_i^q\) comes from the rephasing in the quark fields 
to make the choice of phase convention.
The CP-violating phase \(\delta\) in the representation (3.25) 
is expressed 
with the expression $V$ in Eq.~(\ref{eq-ourckm}) by 
\begin{equation}
\delta =
\mbox{arg}\left[
\left(\frac{V_{12}V_{22}^*}{V_{13}V_{23}^*}\right) + 
\frac{|V_{12}|^2}{1-|V_{13}|^2}
\right] \ ,
\end{equation}
so that we obtain
\begin{equation}
\delta = \pm (80^\circ \pm 22^\circ).
\end{equation}
It is interesting that nearly maximal \(|\sin\delta|\) is realized 
in the present model.\par

The rephasing invariant Jarlskog parameter $J$ \cite{Jarlskog} is 
defined by 
$J={\rm Im}(V_{us}V^*_{cs}V^*_{ub}V_{cb})$. 
In the present model with (3.6)--(3.11), the parameter \(J\) is 
given by
\begin{eqnarray}
J
& =&|\sigma|^2|\rho|c_u s_u c_d s_d \sin \frac{\delta_3+\delta_2}{2} 
\nonumber\\[.1in] 
%& =&\frac{|V_{ub}||V_{cb}||V_{td}||V_{ts}||V_{tb}|}{1-|V_{tb}|^2}\sin 
%\frac{\delta_3+\delta_2}{2}\nonumber\\ 
& =&
\frac{|V_{ub}|}{|V_{cb}|} 
\frac{|V_{td}||V_{ts}||V_{tb}|}{1+|V_{ub}/V_{cb}|^2}
\sin \frac{\delta_3+\delta_2}{2}.
\end{eqnarray}
Using the relation $|V_{td}|\simeq|V_{cb}||V_{us}|$ in (\ref{eq3039}), 
and the
experimental findings $|V_{us}|^2 \gg |V_{cb}|^2 \gg |V_{ub}|^2$, 
$|V_{ts}|\simeq|V_{cb}|$, and $|V_{tb}|\simeq 1$, we obtain 
\begin{equation}
J\simeq|V_{ub}||V_{cb}||V_{us}|\sin\frac{\delta_3+\delta_2}{2}.
\end{equation}
On the other hand, in the standard expression of $V$, (3.25), $J$
is given by
\begin{eqnarray}
J& =&
c_{13}^2 s_{13}c_{12}s_{12}c_{23}s_{23}\sin\delta
\nonumber\\[.1in]
& =&
\frac{|V_{ud}||V_{us}||V_{ub}||V_{cb}||V_{tb}|}{1-|V_{ub}|^2}\sin\delta
\simeq|V_{us}||V_{ub}||V_{cb}|\sin\delta. 
\end{eqnarray}
Comparing Eq. (3.29) with Eq. (3.30), we obtain
\begin{equation}
\sin\delta \simeq \sin\frac{\delta_3+\delta_2}{2}.
\end{equation}
By using the numerical results (3.17)--(3.21), we obtain
\begin{equation}
|J| = (1.91 \pm 0.38)\times 10^{-5}.
\end{equation}

%%%%%%%%%%%%%%%%%%%%%%%%%%%%%%%%%%%%%%%%%%%%%%%%%%%%%%%%%%

%%%%%%%%%%  chap 4  %%%%%%%%%%%%%%
\section{Lepton mixing matrix}

Let us discuss the lepton sectors. 
We assume that the neutrino masses are generated via the seesaw mechanism 
\cite{Yanagida}:
\be
M_\nu=-M_D M_R^{-1} M_D^{T} \ .
\ee
Here $M_D$ and $M_R$ are the Dirac neutrino and the right-handed Majorana 
neutrino mass matrices, which are defined by  $\overline{\nu}_L M_D \nu_R$ 
and $\overline{\nu}_R^c M_R \nu_R$, respectively.
Since $M_D = P_\nu^\dagger \widehat{M}_D P_\nu^\dagger$ and
$M_R = P_\nu^\dagger \widehat{M}_R P_\nu^\dagger$ according to the
assumption (2.9), we obtain
\bea
M_\nu &  =& -P_\nu^\dagger \widehat{M}_D \widehat{M}_R^{-1}
\widehat{M}_D^T P_\nu^\dagger \nonumber \\[.1in]
& =& P_{\nu}^\dagger
\left(
	\begin{array}{ccc}
	0    &  \sqrt{\frac{m_2 m_1}{2}}  &  \sqrt{\frac{m_2 m_1}{2}}\\
	\sqrt{\frac{m_2 m_1}{2}} &  \frac{1}{2}m_3 
\left(1+\frac{m_2-m_1}{m_3}\right) &  -\frac{1}{2}m_3
\left(1-\frac{m_2-m_1}{m_3}\right)\\
	\sqrt{\frac{m_2 m_1}{2}} &  -\frac{1}{2}m_3
\left(1-\frac{m_2-m_1}{m_3}\right) &  \frac{1}{2}m_3 
\left(1+\frac{m_2-m_1}{m_3}\right)
	\end{array}
\right)P_{\nu}^\dagger \ .
\eea
Here and hereafter, \(m_1\), \(m_2\) and \(m_3\) denote neutrino masses 
unless they are specifically mentioned.
In the last expression, we have used the fact\footnote{
The seesaw invariant texture form was discussed systematically in 
\cite{nishiura}.}
that the product of $A B^{-1} A$ of the matrices $A$ and $B$ 
with the texture (1.1) with (1.2) again becomes a matrix
with the texture (1.1) with (1.2).

On the other hand, the charged lepton mass matrix $M_e$ is given by
\bea
M_e  =
P_{e}^\dagger
\left(
	\begin{array}{ccc}
	0    &  \sqrt{\frac{m_\mu m_e}{2}}  &  \sqrt{\frac{m_\mu m_e}{2}}\\
	\sqrt{\frac{m_\mu m_e}{2}} &  \frac{1}{2}m_\tau 
\left(1+\frac{m_\mu-m_e}{m_\tau}\right) &  -\frac{1}{2}m_\tau
\left(1-\frac{m_\mu-m_e}{m_\tau}\right)\\
	\sqrt{\frac{m_\mu m_e}{2}} &  -\frac{1}{2}m_\tau
\left(1-\frac{m_\mu-m_e}{m_\tau}\right) &  \frac{1}{2}m_\tau 
\left(1+\frac{m_\mu-m_e}{m_\tau}\right)
	\end{array}
\right)P_{e}^\dagger \ ,
\eea
where $m_e,~m_\mu$ and $m_\tau$ are charged lepton masses.

Those mass matrices \(M_e\) and \(M_\nu\) are 
diagonalized as $(P_e^\dagger O_e)^\dagger M_e (P_e O_e)=D_e$ and
$(P_\nu^\dagger O_\nu)^\dagger M_\nu (P_\nu O_\nu)=D_\nu$,
respectively, where
\begin{equation}
O_{e}=
\left(
\begin{array}{ccc}
{ c_{e}}&
{ s_{e}}&
{0} \\
{-\frac{s_{e}}{\sqrt{2}}}&
{\frac{c_{e}}{\sqrt{2}}}&
{-\frac{1}{\sqrt{2}}} \\
{-\frac{s_{e}}{\sqrt{2}}}&
{\frac{c_{e}}{\sqrt{2}}}&
{\frac{1}{\sqrt{2}}}
\end{array}
\right), \quad 
O_{\nu}=
\left(
\begin{array}{ccc}
{ c_{\nu}}&
{ s_{\nu}}&
{0} \\
{-\frac{s_{\nu}}{\sqrt{2}}}&
{\frac{c_{\nu}}{\sqrt{2}}}&
{-\frac{1}{\sqrt{2}}} \\
{-\frac{s_{\nu}}{\sqrt{2}}}&
{\frac{c_{\nu}}{\sqrt{2}}}&
{\frac{1}{\sqrt{2}}}
\end{array}
\right). 
\end{equation}
Here \(c_e\) and \(s_e\) are obtained from Eq. (\ref{eq2012}) by 
replacing \(m_1\) and \(m_2\) in it 
by \(m_e\) and \(m_\mu\); 
\(c_\nu\) and \(s_\nu\) are also obtained by taking  the neutrino
masses \(m_i\). 
Therefore, the Maki--Nakagawa--Sakata--Pontecorv (MNSP) lepton mixing matrix 
\cite{MNS} \(U\) can be written as
\begin{eqnarray}
U
& =&O_e^T P O_\nu  \nonumber\\[.1in]
& =&
\left(
\begin{array}{ccc}
c_ec_\nu+\rho_\nu s_e s_\nu \quad & c_e s_\nu-\rho_\nu s_e c_\nu 
\quad & -{\sigma_\nu}s_e \\
s_ec_\nu-\rho_\nu c_e s_\nu \quad & s_e s_\nu+\rho_\nu c_e c_\nu 
\quad & {\sigma_\nu}c_e \\
-{\sigma_\nu}s_\nu \quad & {\sigma_\nu}c_\nu \quad & \rho_\nu \\
\end{array}
\right),\label{MNS}
\end{eqnarray}
where \(P \equiv P_{e} P_\nu^\dagger \equiv \mbox{diag}(e^{i\delta_{\nu1}},
e^{i\delta_{\nu2}},e^{i\delta_{\nu3}})\). Hereafter we take 
\(\delta_{\nu1}=0\) without loss of generality.
\par %%%%%%%%%%%%%%%%%%%%%%%%%%%%%%%%%%%%%%%%%%%%%%%%%%%%%%%%%%%%%%%%%%%%%%%
The explicit forms of absolute magnitudes of the components of \(U\) are
given by expressions similar to (3.4)--(3.12), where $|V_{ij}|$,
$(m_u, m_c, m_t)$, and $(m_d, m_s, m_b)$ are replaced by $|U_{ij}|$, 
$(m_1, m_2, m_3)$, and $(m_e, m_\mu, m_\tau)$, respectively.
It should again be noted that the elements of \(U\) are independent of 
\(m_\tau\) and \(m_3\).
The independent parameters of the unitary matrix \(U\) are 
$\theta_e = \tan^{-1}(m_e/m_\mu)$,
$\theta_\nu = \tan^{-1} (m_1/m_2)$, $\delta_{\nu3}$, and $\delta_{\nu2}$. 
Among them, $\theta_e$ is given by charged-lepton masses of the first and 
second generations. Therefore, the model has the three adjustable parameters  
$\delta_{\nu3}$, $\delta_{\nu2}$, and  $m_1/m_2$ to reproduce 
the experimental values\cite{PDG}. 
\par %%%%%%%%%%%%%%%%%%%%%%%%%%%%%%%%%%%%%%%%%%%%%%%%%%
Let us estimate the values $\theta_\nu$, $\delta_{\nu 3}$ and
$\delta_{\nu 2}$ by fitting the experimental data.
In the following discussions we consider the normal mass hierarchy 
\(\Delta m_{23}^2=m_3^2-m_2^2>0\) for the neutrino mass. 
The case of the inverse mass hierarchy 
\(\Delta m_{23}^2<0\) is quite similar to it.
It follows from the CHOOZ\cite{chooz}, solar\cite{sno},
 and atmospheric neutrino 
experiments\cite{skamioka} that
\begin{equation}
|U_{13}|_{\mbox{\tiny exp}}^2 <  0.03 \ .
\end{equation}
From the global analysis of the SNO solar neutrino experiment\cite{sno},
\begin{eqnarray} 
\Delta m_{12}^2=m_2^2-m_1^2= \Delta m_{\mbox{{\tiny sol}}}^2
=5.0 \times 10^{-5}\, \mbox{eV}^2, \label{mat208301} \\
\tan^2 \theta_{12}=\tan^2 \theta_{\mbox{{\tiny sol}}}=0.34, 
\quad \quad \quad \quad
\label{eq20501}
\end{eqnarray}
with  $\chi^2_{min}/{\rm dof} = 57.0/72$,
for the large mixing angle (LMA) MSW solution.
From the atmospheric neutrino experiment\cite{skamioka}, we also have
\begin{eqnarray}
\Delta m_{23}^2=m_3^2-m_2^2 \simeq \Delta m_{\mbox{{\tiny atm}}}^2
= 2.5 \times 10^{-3}\, \mbox{eV}^2, \\ 
\sin^2 2\theta_{23} \simeq \sin^2 2\theta_{\mbox{{\tiny atm}}}=1.0, 
\quad \quad \quad \quad
\label{mat208302}
\end{eqnarray}
with  $\chi^2_{min}/{\rm dof} = 163.2/170$.
\par
Independently of the parameters $\delta_{\nu3}$ and $\delta_{\nu2}$, 
the model predicts the following two ratios:
\begin{eqnarray}
\frac{\left|U_{13}\right|}{\left|U_{23}\right|}& =&\frac{s_e}{c_e}
=\sqrt{\frac{m_e}{m_\mu}}
=\sqrt{\frac{0.487}{103}}=0.0688, \\
\frac{\left|U_{31}\right|}{\left|U_{32}\right|}& =&\frac{s_\nu}{c_\nu}
=\sqrt{\frac{m_1}{m_2}} \ .
\end{eqnarray}
Here we have used the running charged-lepton mass at \(\mu=m_Z\) 
\cite{Fusaoka}: $m_e(m_Z)=0.48684727 \pm 0.00000014\ \mbox{MeV}$,
and $m_\mu(m_Z)=102.75138 \pm 0.00033\ \mbox{MeV}$.
The neutrino mixing angle \(\theta_{\mbox{{\tiny atm}}}\) 
under the constraint
$ |\Delta m^2_{23}|\gg |\Delta m^2_{12}|$  is given by
\begin{eqnarray}
\sin^2 2 \theta_{\mbox{{\tiny atm}}} 
&\equiv&
4 \left| U_{23} \right|^2 \left|U_{33} \right|^2 \nonumber \\
&=&4\left|\rho_\nu \right|^2 \left|\sigma_\nu\right|^2 c^2_e
= \sin^2(\delta_{\nu3}-\delta_{\nu2})
\sqrt{\frac{m_{\mu}}{m_{\mu}+m_e}} \ .
\end{eqnarray}
The observed fact $\sin^2 2\theta_{\mbox{{\tiny atm}}} 
\simeq 1.0$ highly suggests
$\delta_{\nu3}-\delta_{\nu2}\simeq \pi/2$. 
Hereafter, for simplicity, we take
\begin{equation}
\delta_{\nu3}-\delta_{\nu2}=\frac{\pi}{2} \  .
\end{equation}
Under the constraint (4.13), the model predicts 
\begin{equation}
|U_{13}|^2={\frac{1}{2}}{\frac{m_e}{m_\mu+m_e}}=0.00236 \ ,
\ \mbox{or}~~\mbox{sin}^22\theta_{13}=0.00942 \ .
\end{equation}
This value is consistent with the present experimental constraints 
(4.6) and can be checked in neutrino factories 
\cite{cervera}, which have sensitivity to sin$^22\theta_{13}$ for
\begin{equation}
\mbox{sin}^22\theta_{13}\ge 10^{-5}.
\end{equation}

\par
The mixing angle \(\theta_{\mbox{{\tiny sol}}}\) in the present model 
is given by
\begin{eqnarray}
\sin^2 2\theta_{\mbox{{\tiny sol}}}
&\equiv&
4 \left|U_{11}\right|^2 \left|U_{12}\right|^2 \nonumber \\
&\simeq&
\frac{4m_2 m_1}{(m_2+m_1)^2}
\left[1 -\sqrt{2}\cos\frac{\delta_{\nu3}+\delta_{\nu2}}{2}
 \sqrt{\frac{m_e m_2}{m_\mu m_1}}
+ \frac{1}{2}\left( \frac{m_e m_2}{m_\mu m_1}\right)\right]\nonumber \\
&\simeq &\frac{4m_1/m_2}{(1+m_1/m_2)^2} \ , 
\end{eqnarray} 
which leads to
\begin{equation}
\frac{m_1}{m_2} \simeq \tan^2\theta_{\mbox{{\tiny sol}}}=0.34 \ ,
\label{ratio}
\end{equation}
where we have used the best fit value (4.8).
This value (\ref{ratio}) guarantees the validity of the approximation
(4.17), because of $\sqrt{(m_e/m_\mu)/(m_1/m_2)}\simeq 0.12$.
Then, we can obtain the neutrino masses
\begin{eqnarray}
m_1 & = & 0.0026 \, {\rm eV} \ ,\nonumber \\
m_2 & = & 0.0075 \, {\rm eV} \ , \\
m_3 & = & 0.050 \, {\rm eV} \ ,\nonumber
\end{eqnarray}
where we have used the observed best fit values of 
$\Delta m^2_{\mbox{{\tiny sol}}}$
and $\Delta m^2_{\mbox{{\tiny atm}}}$, (4.7) and (4.9), respectively.

%%%%%%%%%%%%%%%%%%%%%%%%%%%%%%%%%%%%%%%%%%%%%%%%%%%%%%%%%%%%%%%%%%%%%%%%%%

Next let us discuss the CP violation phases in the lepton mixing matrix.
The Majorana neutrino fields do not have the freedom of rephasing 
invariance, so that we can use only the rephasing freedom of $M_e$ 
to transform Eq. (\ref{MNS}) to the standard form
\begin{equation}
U_{\rm std} \equiv
\left(
\begin{array}{ccc}
c_{\nu13}c_{\nu12} & c_{\nu13}s_{\nu12}e^{i\beta} & 
s_{\nu13}e^{i(\gamma-\delta_{\nu})} \\
(-c_{\nu23}s_{\nu12}-s_{\nu23}c_{\nu23}s_{\nu13} e^{i\delta_{\nu}})e^{-i\beta}
&c_{\nu23}c_{\nu12}-s_{\nu23}s_{\nu12}s_{\nu13} e^{i\delta_{\nu}} 
&s_{\nu23}c_{\nu13}e^{i(\gamma-\beta)} \\
(s_{\nu23}s_{\nu12}-c_{\nu23}c_{\nu12}s_{\nu13} e^{i\delta_{\nu}})e^{-i\gamma}
 & (-s_{\nu23}c_{\nu12}-c_{\nu23}s_{\nu12}s_{\nu13} 
e^{i\delta_{\nu}})e^{-i(\gamma-\beta)} 
& c_{\nu23}c_{\nu13}  \\
\end{array}
\right) \ ,
\label{majorana}
\end{equation}
as 
\begin{equation}
U_{\rm std} = \mbox{diag}(e^{i\alpha_1^e},e^{i\alpha_2^e},e^{i\alpha_2^e}) 
\ U \
\mbox{diag}(e^{\pm i\pi/2},1,1) \ .
\end{equation}
Here, \(\alpha_i^e\) comes from the rephasing in the charged lepton fields 
to make the choice of phase convention, and 
the specific phase \({\pm \pi/2}\) is added on the right-hand side of \(U\) 
in order to change the neutrino eigen-mass \(m_1\) to a positive quantity.
Similarly to the quark sector, the CP-violating phase \(\delta_{\nu}\) 
in the representation (4.20) is expressed as
\begin{equation}
\delta_\nu = \mbox{arg}
          \left[
             \frac{U_{12}U_{22}^*}{U_{13}U_{23}^*} + 
             \frac{|U_{12}|^2}{1-|U_{13}|^2}
          \right]
 \simeq \mbox{arg} \left(
          \frac{U_{12} U_{22}^*}{U_{13} U_{23}^*} \right)
  \simeq \mbox{arg}\rho_\nu^* +\pi = -\frac{\delta_{\nu 3}
+\delta_{\nu2}}{2} +\pi \ .
\end{equation}
Though the lepton mixing matrix includes the additional Majorana 
phase factors $\beta$ and $\gamma$ \cite{bilenky,Doi}, the number of 
parameters which will become experimentally available in the near 
future is practically four, as in the Dirac case.
The additional phase parameters are determined as 
\begin{equation}
\beta= \mbox{arg} \left( \frac{U_{{\rm std} \ 12}}{U_{{\rm std} \ 11}} \right)
     = \mbox{arg} \left( \frac{U_{12}}{U_{11} e^{\pm i\pi}} \right) 
  \simeq 0 \mp \frac{\pi}{2} \ ,
\end{equation}
and
\begin{equation}
\gamma = \mbox{arg} \left(\frac{U_{{\rm std} \ 13}}{U_{{\rm std} \ 11} } 
e^{i \delta_\nu}\right)\\
       = \mbox{arg} \left(\frac{U_{13}}{U_{11}e^{\pm i\pi}} 
e^{i \delta_\nu}\right)
       \simeq \mbox{arg}(-\sigma_\nu) +\delta_\nu \mp \frac{\pi}{2}
         \simeq \frac{\pi}{2} \mp \frac{\pi}{2} \ ,
\end{equation}
by using the relations \(m_e \ll m_\mu \) and 
\( (\delta_{\nu3}-\delta_{\nu2})/2 \simeq  \pi/4\).
Hence, we can also predict the averaged neutrino mass 
$\langle m_\nu \rangle$ \cite{Doi}, which appears in the neutrinoless 
double beta decay, as follows:
\begin{eqnarray}
\langle m_\nu \rangle & \equiv & \left| -m_1 U_{11}^2 +m_2 U_{12}^2
+m_3 U_{13}^2 \right| \nonumber \\
 & = & \left|-2 \rho_\nu c_e s_e \sqrt{m_1 m_2} 
+ \rho_\nu^2 s_e^2 (m_2-m_1) +m_3 s_e^2 \right| \ .
\end{eqnarray}
The value of (4.25) is highly sensitive to the value of
$(\delta_{\nu 3}+\delta_{\nu 2})/2$, which is unknown at present,
because the values $s_e/c_e = \sqrt{m_e/m_\mu}\simeq 0.070$
and $\sqrt{m_1 m_2}/m_3 \simeq 0.088$ are in the same order.
For $(\delta_{\nu 3}+\delta_{\nu 2})/2=0$, $\pi/2$ and $\pi$,
we obtain the numerical results  $\langle m_\nu \rangle
=$ $0.00018$ eV, $0.00049$ eV and $0.00069$ eV, respectively.
However, these values should not be taken strictly because the
value $m_1/m_2$ is also sensitive to the observed value of
$\tan^2 \theta_{sol}$. In any cases, the predicted value of
$\langle m_\nu \rangle$ will be less than the order of 
$10^{-3}$ eV.

The rephasing-invariant parameter $J$ in the lepton 
sector is defined by $J={\rm Im}(U_{12}U^*_{22}U^*_{13}U_{23})$, 
which is explicitly given by
\begin{eqnarray}
J
&=&|\sigma_{\nu}|^2|\rho_{\nu}|c_{\nu} s_{\nu} c_{e} s_{e} 
\sin{\frac{\delta_{\nu3}+\delta_{\nu2}}{2}} \nonumber\\ 
%& =&\frac{|U_{13}||U_{23}||U_{31}||U_{32}||U_{33}|}{1-|U_{33}|^2} 
%\sin{\frac{\delta_{\nu3}+\delta_{\nu2}}{2}} \nonumber\\ 
&=&
\frac{|U_{13}|}{|U_{23}|} 
\frac{|U_{31}||U_{32}||U_{33}|}{1+|U_{13}/U_{23}|^2}
\sin \frac{\delta_{\nu3}+\delta_{\nu2}}{2} \le 
\frac{|U_{13}|}{|U_{23}|} 
\frac{|U_{31}||U_{32}||U_{33}|}{1+|U_{13}/U_{23}|^2} \ .
\end{eqnarray}
The upper bound is described in terms of the ratio ${m_1}/{m_2}$,
so that we obtain 
\begin{equation}
J \leq 0.019.
\end{equation}
It should be noted that if we again assume the maximal CP
violation in the lepton sector, the magnitude of the
rephasing invariant $|J|$ can be considerably larger than 
in the quark sector, $|J_{\rm quark}|\simeq 2 \times 10^{-5}$.

%%%%%%%%%%  chap 6  %%%%%%%%%%%%%%
\section{conclusion}

In conclusion, stimulated by recent neutrino data, which suggest
a nearly bimaximal mixing, we have investigated a possibility
that all the mass matrices of quarks and leptons have the same
texture as the neutrino mass matrix.  
We have assumed that the mass matrix form is constrained by 
a discrete symmetry Z$_3$ and a permutation symmetry S$_2$, i.e.
that the texture is given by the form (1.1) with (1.2).
The most important feature of the present model is that
the textures (1.1)--(1.2) are practically applicable to 
the predictions
at the low energy scale (the electroweak scale), although
we assume that the textures are exactly given at a unification
scale.

It is well known that the matrix form (1.1) leads to a bimaximal mixing
in the neutrino sector.
In the present model, the mixing angle $\theta^f_{12}$ between
the first and second generations is given by 
$$
\tan \theta^f_{12} = \sqrt{m^f_1/m^f_2} \ ,
\eqno(5.1)
$$
where $m^f_1$ and $m^f_2$ are the first and second generation fermion
masses.
This leads to a large mixing in the lepton mixing matrix (MNSP matrix)
$U$ with $m_1 \sim m_2$ (neglecting $\tan \theta_{12}^e 
= \sqrt{m_e/m_\mu}$
in the charge lepton sector), and it also leads to the famous 
formula \cite{Vus} $|V_{us}| \simeq \sqrt{m_d/m_s}$ in the
quark mixing matrix (CKM matrix) $V$ (neglecting 
$\tan \theta_{12}^u = \sqrt{m_u/m_c}$ in the up-quark sector).
In the present model the mixing angle $\theta^f_{23}$ between 
the second and third generation is fixed as $\theta^f_{23}=\pi/4$.
However, the (2,3) component of the quark mixing matrix $V$ 
(and also the lepton mixing matrix  $U$) is highly dependent on 
the phase difference $\delta_3-\delta_2$, as follows
$$
V_{23} = \frac{1}{\sqrt{1+m^u_1/m^u_2}}
\sin\frac{\delta_3-\delta_2}{2} \ ,
\eqno(5.2)
$$
where $\delta_i = \delta^u_i -\delta^d_i$.
Replacing the arguments by their leptonic counterparts,
we have the same form for $U_{23}$.
We have understood the observed values $V_{23}$ and
$U_{23}$ by taking $(\delta_3-\delta_2)/2$ as a small 
value for the quark sectors and as $\pi/2$ for the lepton sectors,
respectively.
As predictions, which are independent of such phase
parameters, there are two relations
$$
\frac{|V_{ub}|}{|V_{cb}|}=\sqrt{\frac{m_u}{m_c}} \ , \ \ 
\frac{|V_{td}|}{|V_{ts}|}=\sqrt{\frac{m_d}{m_s}} \ , 
\eqno(5.3)
$$
(and the similar relations for $U$).
The relations (5.3) are in good agreement with
experiments.
The relation $|U_{13}/U_{23}|=\sqrt{m_e/m_\mu}$
in the lepton sectors leads to $|U_{13}|^2
\simeq m_e/2m_\mu = 0.0024$ if we accept
$\sin^2 2\theta_{\rm atm}=1.0$.
This value will be testable in the near future.

Since, in the present model, each mass matrix
$M_f$ (i.e. the Yukawa coupling $Y_f$) takes
different values of $A_f$, $B_f$, and so on,
the present model cannot be embedded into a GUT
scenario.
In spite of such a demerit, however, it is worth 
while noting that it can give 
a unified description of quark and
lepton mass matrices with the same texture.

%%%%%%%%%%%%%%%%%%%%%%%%%%%%%%%%%%%%%%%%%%%%%%%%%%%%%%%%%%%%%
\vspace{5mm}
\centerline{\bf ACKNOWLEDGEMENT}

One of the authors (YK) wishes to acknowledge the
hospitality of the Theory group at CERN, where this
work was completed.
YK also thank Z.~Z.~Xing for informing many helpful
references.

%%%%%%%%%%%%%%%%%%%%%%%%%%%%%%%%%%%%%%%%%%%%%%%%%%%%%%%%
\vspace{5mm}
\centerline{\bf APPENDIX}

The mass matrix texture (1.1) with (1.2), which is defined at
the unification energy scale $\mu=M_X$, is applicable to
the phenomenology at the electroweak scale $\mu=m_Z$.
In the present appendix, we demonstrate this for the 
quark mass matrices $M_u$ and $M_d$.

It is well known \cite{evol} that the energy scale dependences
$R(A)=A(\mu)/A(M_X)$ for observable quantities $A$
approximately satisfy  the relations
$R(|V_{ub}|) \simeq R(|V_{cb}|) \simeq R(|V_{td}|) 
\simeq R(|V_{ts}|) \simeq R(m_d/m_b) \simeq R(m_s/m_b)$, 
and that  the ratios $R(|V_{us}|)$,  $R(|V_{cd}|)$,
$R(m_d/m_s)$ and $R(m_u/m_c)$ are approximately constant.
This is caused by the fact that the Yukawa coupling constant
$y_t$ of the top quark is extremely large with respect to 
other coupling constants.
The above relations on $R$ are well explained by the
approximation $y_t^2 \gg y_b^2, y_c^2, \cdots$.
Therefore, we will also use approximation below.

The one-loop RGE for the Yukawa coupling constants $Y_f$ 
($f=u,d$) has the form
$$
\frac{dY_f}{dt} = \frac{1}{16\pi^2}\left(
C_f {\bf 1} + C_{ff} Y_f Y_f^\dagger
+C_{ff'} Y_{f'} Y_{f'}^\dagger \right) Y_f \ ,
\eqno(A.1)
$$
where $f'=d$ ($f'=u$) for $f=u$ ($f=d$), and
the coefficients $C_f$, $C_{ff}$ and $C_{ff'}$ are
energy scale dependent factors which are calculated
from the one-loop Feynman diagrams.
We start from the Yukawa coupling constants $Y_f(M_X)$,
corresponding to the mass matrix form (1.1) [with (1.2)].

Since the matrix $Y_u Y_u^\dagger$ is approximately given
by
$$
Y_u(M_X) Y_u^\dagger (M_X) \simeq \frac{m_t^2 }{v_u^2}
\left(
\begin{array}{ccc}
0 & 0 & 0 \\
0 & 1 & -e^{+i \delta_u} \\
0 & -e^{-i \delta_u} & 1
\end{array} \right) \ ,
\eqno(A.2)
$$
where $\delta_u = \delta_3^u -\delta_2^u$, 
$v_u/\sqrt{2}=\langle H_u^0 \rangle$, and 
we have used the relations (1.6) and the approximation
$y_t^2 \gg y_b^2, y_c^2, \cdots$,
the up-quark Yukawa coupling constant $Y_u(\mu)$ in the 
neighbourhood of $\mu = M_X$ is given by the form
$$
Y_u (\mu) \simeq r_u(\mu) \left[ {\bf 1} +
\varepsilon_u(\mu) 
\left(
\begin{array}{ccc}
0 & 0 & 0 \\
0 & 1 & -e^{+i \delta_u} \\
0 & -e^{-i \delta_u} & 1
\end{array} \right) \right] Y_u (M_X) 
\simeq \frac{r_u(\mu)}{v_u/\sqrt{2}} 
$$
$$
%\simeq \frac{r_u(\mu)}{v_u/\sqrt{2}} 
\left(
\begin{array}{ccc}
0 & A_u e^{-i \delta_2^u} & A_u e^{-i\delta_3^u} \\
A_u (1+\varepsilon_u -\varepsilon_u) e^{-i \delta_2^u} & 
B_u (1+\varepsilon_u -\varepsilon_u C_u/B_u) 
e^{-2i \delta_2^u} & C_u(1+\varepsilon_u -\varepsilon_u B_u/C_u)
e^{-i (\delta_2^u+\delta_3^u)} \\
A_u (1+\varepsilon_u -\varepsilon_u) e^{-i \delta_3^u} & 
C_u(1+\varepsilon_u -\varepsilon_u B_u/C_u)
e^{-i (\delta_2^u+\delta_3^u)}  & 
B_u (1+\varepsilon_u -\varepsilon_u C_u/B_u) 
e^{-2i \delta_3^u}
\end{array} \right)
\ .
\eqno(A.3)
$$
Although this form is one in $\mu \simeq M_X$, but, since the texture
keeps the same form under the small change of energy scale, 
as a result, the texture of $Y_u(\mu)$ given by (A.3) holds
at any energy scale $\mu$.
Therefore, we can obtain the expression (1.1) at an arbitrary
energy scale $\mu$.
(The demonstration (A.3) has been done
for the case $P_R=P_L^\dagger$ mentioned in (2.10). 
However, the conclusion does not depend on this choice.)

On the other hand, the evolution of the down-quark Yukawa
coupling constant $Y_d(\mu)$ is somewhat complicated.
By a way similar to (A.3), we obtain
$$
Y_d (\mu) \simeq r_d(\mu)  
\left(
\begin{array}{ccc}
1 & 0 & 0 \\
0 & 1+\varepsilon_d & -\varepsilon_d e^{+i \delta_u} \\
0 & -\varepsilon_d e^{-i \delta_u} & 1+\varepsilon_d
\end{array} \right)  Y_d (M_X) 
\simeq \frac{r_d(\mu)}{v_d/\sqrt{2}} 
$$
$$
%\simeq \frac{r_d(\mu)}{v_d/\sqrt{2}} 
P_d^\dagger \left(
\begin{array}{ccc}
0 & A_d  & A_d  \\
A_d (1+\varepsilon_d -\varepsilon_d e^{+i (\delta_u-\delta_d)}) & 
B_d (1+\varepsilon_d -\varepsilon_d e^{+i(\delta_u-\delta_d)} C_d/B_d) 
& C_d(1+\varepsilon_d -\varepsilon_d e^{+i(\delta_u-\delta_d)} B_d/C_d)
 \\
A_d (1+\varepsilon_d -\varepsilon_d e^{-i (\delta_u-\delta_d)}) & 
C_d(1+\varepsilon_d -\varepsilon_d e^{-i(\delta_u-\delta_d)} B_d/C_d)
 & B_d (1+\varepsilon_d -\varepsilon_d e^{-i(\delta_u-\delta_d)} C_d/B_d) 
\end{array} \right) P_d^\dagger
\ ,
\eqno(A.4)
$$
where $\delta_u -\delta_d=(\delta_3^u-\delta_2^u)-(\delta_3^d-\delta_2^d)
=\delta_3-\delta_2$.
Note that the part that is sandwiched between $P_d^\dagger$ and $P_d^\dagger$
includes imaginary parts and those phase factors cannot be removed
by an additional phase matrix $P_d(\mu)$ into the form $P_d^\dagger(\mu)
\widehat{Y}_d(\mu) P_d^\dagger (\mu)$.
However, the quantity that has the physical meaning is 
$Y_d Y_d^\dagger$.
When we define
$$
\xi e^{-i\alpha} =1 +\varepsilon_d (1-e^{-i(\delta_u-\delta_d)}) \ , \ \  \
\eta e^{+i\beta} =1 +\varepsilon_d (1+e^{+i(\delta_u-\delta_d)}) \ , 
\eqno(A.5)
$$
we obtain
$$
Y_d(\mu) Y_d^\dagger(\mu) \simeq  \frac{r_d^2(\mu)}{v_d^2/2} 
P_d^\dagger P_\beta 
\left(
\begin{array}{ccc}
A_d^2 & A_d (B_d+C_d) & A_d(B_d+C_d)  \\
A_d (B_d+C_d)\eta & A_d^2\xi^2 +(B_d^2+C_d^2)\eta^2 
& A_d^2\xi^2 e^{2i(\alpha-\beta)}+2 B_d C_d\eta^2  \\
A_d (B_d+C_d)\eta & A_d^2\xi^2 e^{2i(\alpha+\beta)}+2 B_d C_d\eta^2 
 &  A_d^2\xi^2 +(B_d^2+C_d^2)\eta^2 
\end{array} \right) P_\beta P_d^\dagger
\ ,
\eqno(A.6)
$$
where
$$
P_\beta ={\rm diag}(1, e^{i\beta}, e^{-i\beta}) \ ,
\eqno(A.7)
$$
so that we can obtain a real matrix for the part which is sandwiched
by the phase matrix $P_d P_\beta$ under the approximation
$A_d^2/|B_d C_d| \simeq 0$.
This means that we can practically write
$$
\widehat{Y}_d(\mu) \simeq \frac{r_d(\mu)}{v_d/\sqrt{2}}
\left( 
\begin{array}{ccc}
0 & A_d & A_d \\
A_d \xi & B_d \eta & C_d \eta \\
A_d \xi & C_d \eta & B_d \eta \\
\end{array} \right) \ ,
\eqno(A.8)
$$
with
$$
P_d^\dagger (\mu) = {\rm diag}(1, e^{-i(\delta_2^d-\beta)},
e^{-i(\delta_3^d+\beta)}) \ ,
\eqno(A.9)
$$
at an arbitrary energy scale $\mu$.
It should be noted that the changes of the phases
$\delta_2^d \rightarrow \delta_2^d -\beta$ and
$\delta_3^d \rightarrow \delta_3^d +\beta$ do not
come from the evolution of the phases $\delta_2^d (\mu)$
and $\delta_3^d (\mu)$, but they are brought effectively by 
absorbing the unfactorizable phase parts in $Y_d(\mu)$.
Thus, we can again use the texture (1.1) at an arbitrary
energy scale $\mu$ from a practical point of view.

In the Yukawa coupling constants $Y_e$ and $Y_\nu$ of
the leptons, the RGE effects are not so large as in 
the quark sectors.
In the charged lepton sector, since $m_\tau^2 \gg m_\mu^2
\gg m_e^2$, we can again demonstrate that the expression
(1.1) is applicable at an arbitrary energy scale 
in a way similar to the quark sectors.
For the neutrino Yukawa coupling constant $Y_\nu(\mu)$,
the evolution equation is different from (A.1). 
We must use the RGE for the seesaw operator \cite{evol-nu}.
However, the calculation and result are essentially the same
as those in $Y_u(\mu)$, $Y_d(\mu)$ and $Y_e(\mu)$, 
because $m_3^2\gg m_2^2 >m_1^2$ in the present model.

Finally, we would like to add that these conclusions 
on the evolution of the mass matrices $M_f$ ($f=u,d,e, \nu$)
are exactly confirmed by numerical study, without 
approximation.

%%%%%%%%%%%%%%%%%%%%%%% ref %%%%%%%%%%%%%%%%%%%%%%%%%%%%%%%%%%%%%%%%%%%

%%%%%%%%%%%%%%%%%%%%%%%%%%%%%%%%%%%%%%%%%%%%%%%%%%%%%%%%%%%%%%%


\begin{thebibliography}{99}
\bibitem{skamioka}
M. Shiozawa, talk at Neutrino 2002
 (http://neutrino.t30.physik.tu-muenchen.de/).
\bibitem{Fukuyama}
T. Fukuyama and H. Nishiura,
hep-ph/9702253; in Proceedings of the International 
Workshop on Masses and 
Mixings of Quarks and Leptons,  Shizuoka, Japan, 1997, 
edited by Y.~Koide (World Scientific, Singapore, 1998), p. 252.
\bibitem{Ma}
E. Ma and  M. Raidal, Phys. Rev. Lett. {\bf 87}, 011802 (2001).
\bibitem{Lam}
C.S. Lam, Phys. Lett. {\bf B507}, 214 (2001).
\bibitem{Balaji}
K.R.S. Balaji, W. Grimus and T. Schwetz, 
Phys. Lett. {\bf B508}, 301 (2001).
\bibitem{Grimus}
W. Grimus and L. Lavoura, Acta Phys. Pol. {\bf B32}, 3719 (2001).
\bibitem{Nishiura}
H. Nishiura, K. Matsuda, T. Kikuchi and T. Fukuyama, 
Phys. Rev. {\bf D65}, 097301 (2002).
%
\bibitem{howmanyH} For instance,
K.~Oda, E.~Takasugi, M.~Tanaka and M.~Yoshimura, Phys.~Rev.
{\bf D59}, 055001 (1999);
K.~Matsuda, Y.~Koide and T.~Fukuyama, Phys.~Rev.
{\bf D64}, 053015 (2001).
%
\bibitem{Fuk-Okd}
For instance, T.~Fukuyama and N.~Okada, hep-ph/0205066.
%
\bibitem{CKM}
N. Cabibbo, Phys. Rev. Lett. {\bf 10}, 531 (1963);
M. Kobayashi and T. Maskawa, Prog. Theor. Phys. {\bf 49}, 652 (1973).
\bibitem{PDG}
Particle Data Group, K.~Hagiwara {\it et al.}, 
Phys.~Rev. {\bf D66}, 010001 (2002).
\bibitem{Fusaoka}
H. Fusaoka and Y. Koide, Phys. Rev. {\bf D57}, 3986 (1998).
%
\bibitem{Branco}
G.~C.~Branco and L.~Lavoura, Phys.~Rev. {\bf D44}, R582 (1991).
%
\bibitem{Fritzsch}
H.~Fritzsch and Z.~Z.~Xing, Phys.~Lett. {\bf B413}, 396 (1997);
Phys.~Rev. {\bf D57}, 594 (1998).
For the application of their ansatz to the lepton sectors, see
H.~Fritzsch and Z.~Z.~Xing, Phys.~Lett. {\bf B440}, 313 (1998);
Phys.~Rev. {\bf D61}, 073016 (2000); Z.~Z.~Xing, Nucl.~Phys. B
(Proc.~Suppl.) {\bf 85}, 187 (2000).
%
\bibitem{Jarlskog}
C.~Jarlskog, Phys.~Rev.~Lett. {\bf 55}, 1839 (1985);
O.~W.~Greenberg,  Phys.~Rev. {\bf D32}, 1841 (1985);
I.~Dunietz, O.~W.~Greenberg and D.-d.~Wu,  Phys.~Rev.~Lett. {\bf 55}, 
2935 (1985);
C.~Hamzaoui and A.~Barroso,  Phys.~Rev. {\bf D33}, 860 (1986).
%
\bibitem{Yanagida}
T. Yanagida, in Proceedings of the Workshop 
on the Unified Theory and Baryon Number in the Universe, 
edited by O. Sawada and A. Sugamoto (KEK, Tsukuba, 1979), p.~95; 
M.~Gell-Mann, P.~Ramond and R.~Slansky, in 
Supergravity, edited by P. van Nieuwenhuizen and D. Freedman 
(North-Holland, Amsterdam,1979), p. 315;
G.~Senjanovi{\'c} and R.~N.~Mohapatra, Phys.~Rev.~Lett. {\bf 44}, 912 (1980).
\bibitem{nishiura}
H.~Nishiura, K.~Matsuda and T.~Fukuyama, Phys.~Rev. {\bf D60}, 013006 (1999).
%
\bibitem{MNS} Z.~Maki, M.~Nakagawa and S.~Sakata,
Prog.~Theor.~Phys. {\bf 28}, 870 (1962); 
B.~Pontecorvo, Zh.~Eksp.~Theor.~Fiz.
{\bf 33}, 549 (1957);
Sov.~Phys.~JETP {\bf 26}, 984 (1968).
%
\bibitem{chooz}
M. Apollonio {\it et al.}, 
Phys.~Lett. {\bf B466}, 415 (1999).
\bibitem{sno}
Q.~R.~Ahmad {\it et al.}, 
Phys.~Rev.~Lett. {\bf 89}, 011301 and  011302 (2002). 
%
\bibitem{cervera}
A.~Cervera {\it et al.}, Nucl.~Phys. {\bf B579}, 17 (2000);
Erratum, {\it ibid.} {\bf B593}, 731 (2000).
%
\bibitem{bilenky}
S.~M.~Bilenky, J.~Hosek and S.~T.~Petcov, Phys.~Lett. {\bf 94B}, 495 (1980);
J.~Schechter and J.~W.~F.~Valle, Phys.~Rev. {\bf D22}, 2227 (1980);
 A.~Barroso and J.~Maalampi, Phys.~Lett. 
{\bf 132B}, 355 (1983).
%
\bibitem{Doi} 
M.~Doi, T.~Kotani, H.~Nishiura, K.~Okuda and E.~Takasugi, Phys.~Lett. 
{\bf 102B}, 323 (1981).
%
\bibitem{Vus}
S.~Weinberg, Ann.~N.Y.~Acad.~Sci. {\bf 38}, 185 (1977);
H.~Fritzsch, Phys.~Lett. {\bf 73B}, 317 (1978); 
Nucl.~Phys. {\bf B155}, 189 (1979);
H.~Georgi and D.~V.~Nanopoulos, {\it ibid.} {\bf B155}, 52 (1979).
%
\bibitem{evol}
%B.~Grzadkowski and M.~Lindner, Phys.~Lett. {\bf B193},
%71 (1987); B.~Grzadkowski, M.~Lindner and S.~Theisen, 
%Phys.~Lett. {\bf B198}, 64 (1987);
For example, see M.~Olechowski and S.~Pokorski, Phys.~Lett. {\bf B257},
288 (1991).
For a recent study, see for einstance S.~R.~Ju\'{a}rez W., S.~F.~Herrera H.,
P.~Kielanowski and G.~Mora, hep-ph/0206243.
%
\bibitem{evol-nu} 
P.~H.~Chankowski and Z.~Pluciennik, Phys.~Lett. {\bf B316},
312 (1993);
K.~S.~Babu, C.~N.~Leung and J.~Pantaleone, {\it ibid.}
{\bf B319}, 191 (1993).
\end{thebibliography}
\end{document}